\documentclass[journal]{IEEEtran}
\usepackage{amsmath,graphicx,url,times,hyperref}
\usepackage{amssymb}
\usepackage{mathrsfs}
\usepackage{booktabs}
\usepackage[T1]{fontenc}
\usepackage{algorithm}
\usepackage{algpseudocode}
\usepackage{textcomp}

\usepackage[usenames, dvipsnames]{color}
\usepackage{booktabs}

\newcommand{\qk}[1] {{\color{black} #1}}

\usepackage{multirow}
\ifCLASSINFOpdf
\else
\fi
\begin{document}
%
\title{{\huge High-resolution Piano Transcription with Pedals by Regressing Onset and Offset Times}}
%
%
%


\author{Qiuqiang Kong, Bochen Li, Xuchen Song, Yuan Wan, Yuxuan Wang\thanks{Q. Kong, B. Li, X. Song, Y. Wan, Y. Wang are with ByteDance. (e-mail: kongqiuqiang@bytedance.com; bochenli@bytedance.com; xuchen.song@bytedance.com, wanyuan.0626@bytedance.com, wangyuxuan.11@bytedance.com). \textit{(Qiuqiang Kong is first and corresponding author.)}}}

\maketitle

\begin{abstract}
Automatic music transcription (AMT) is the task of transcribing audio recordings into symbolic representations. Recently, neural network-based methods have been applied to AMT, and have achieved state-of-the-art results. However, many previous systems only detect the onset and offset of notes frame-wise, so the transcription resolution is limited to the frame hop size. There is a lack of research on using different strategies to encode onset and offset targets for training. In addition, previous AMT systems are sensitive to the misaligned onset and offset labels of audio recordings. Furthermore, there are limited researches on sustain pedal transcription on large-scale datasets. In this article, we propose a high-resolution AMT system trained by regressing precise onset and offset times of piano notes. At inference, we propose an algorithm to analytically calculate the precise onset and offset times of piano notes and pedal events. We show that our AMT system is robust to the misaligned onset and offset labels compared to previous systems. Our proposed system achieves an onset F1 of 96.72\% on the MAESTRO dataset, outperforming previous onsets and frames system of 94.80\%. Our system achieves a pedal onset F1 score of 91.86\%, which is the first benchmark result on the MAESTRO dataset. We have released the source code and checkpoints of our work at \url{https://github.com/bytedance/piano\_transcription}.
\end{abstract}


\section{Introduction}\label{sec:introduction}

Automatic music transcription (AMT) \cite{raphael2002automatic, redmon2016you, benetos2018automatic} is the task of transcribing audio recordings into symbolic representations \cite{benetos2013automatic}, such as piano rolls, guitar fretboard charts and Musical Instrument Digital Interface (MIDI) files. AMT is an important topic of music information retrieval (MIR) and is a bridge between audio-based and symbolic-based music understanding. AMT systems have several applications, such as score following \cite{li2016approach}, audio to score alignment \cite{niedermayer2010multi}, and score-informed source separation \cite{duan2011soundprism}. In industry, AMT systems can be used to create music education software for music learners. For music production, AMT can be used to transcribe audio recordings into MIDI files for intelligent music editing. AMT systems can also be used for symbolic-based music information retrieval and can be used to analyze unarchived music, such as jazz improvisations.

Piano transcription is an essential task of AMT. The task is to transcribe piano solo recordings into music note events with pitch, onset, offset, and velocity. Piano transcription is a challenging task due to the high polyphony of music pieces. Early works of piano transcription \cite{scheirer1995using, dixon2000computer, marolt2000transcription} include using discriminative models, such as support vector machines to predict the presence or absence of notes in audio frames \cite{poliner2006discriminative}. To address the multiple pitch estimation problem, a probabilistic spectral smoothness principle was proposed in \cite{emiya2009multipitch} for piano transcription. A combination of frequency domain and time domain method was proposed for piano transcription in \cite{bello2006automatic}, where authors assumed that signals are a linearly weighted sum of waveforms in a database of individual piano notes. Non-negative matrix factorizations (NMFs) and non-negative sparse codings \cite{abdallah2004polyphonic} have been proposed to decompose spectrogram into polyphonic notes \cite{o2014polyphonic}, where signals are decomposed into the multiplication of dictionaries and activations for transcribing polyphonic music. To model different onset and decay states of piano notes, an attack and decay system was proposed in \cite{cheng2016attack}. Other AMT systems include using unsupervised learning method \cite{berg2014unsupervised}, connectionist approaches \cite{marolt2004connectionist} and fast convolutional sparse coding methods \cite{cogliati2015piano}. 

Recently, neural networks have been applied to tackle the ATM problem. A deep belief network was proposed to learn feature representations for music transcription in \cite{nam2011classification}. Fully connected neural networks, convolutional neural networks (CNNs), and recurrent neural networks (RNNs) \cite{bock2012polyphonic, sigtia2016endtoend, kelz2016potential, kelz2019multitask} were proposed to learn regressions from audio input to labelled ground truths. Recently, onsets and frames systems \cite{hawthorne2017onsets, hawthorne2018enabling} were proposed to predict both onsets and frame-wise pitches of notes and have achieved state-of-the-art results in piano transcription. To improve the encoding of onset targets, several works including SoftLoc \cite{schroeter2019softloc} and non-discrete annotations \cite{elowsson2018polyphonic, elowsson2018modeling, gkiokas2017convolutional} have been proposed.

However, there are several limitations of previous AMT methods \cite{hawthorne2017onsets, hawthorne2018enabling}. First, those previous works used binarized values to encode onsets and offsets in each frame. Therefore, the transcription resolution in the time domain is limited to the frame hop size between adjacent frames. For example, a hop size of 32 ms was used in \cite{hawthorne2017onsets}. So that the transcription resolution is limited to 32 ms. In many scenarios, high-resolution transcription systems can be useful for music analysis. For example, Gobel \cite{goebl2001melody, goebl2003role} analyzed the melody lead phenomenon in milliseconds. The second limitation is that the modeling of onsets and offsets can be improved. In \cite{hawthorne2017onsets}, authors empirically used an onset length of 32 ms and claim that almost all onsets will end up spanning exactly two frames. However, there is a lack of explanation of labelling two frames as onsets performs better than other representations. The analysis from \cite{cheng2016attack} shows that the attack of a piano note can last for several frames instead of only one frame. A note can be modeled in a more natural way with an attack, decay, sustain and release (ADSR) states \cite{kelz2019deep}. In addition, the targets in \cite{hawthorne2017onsets} are sensitive to the misalignment between audio recordings and labels. For example, if onset is misaligned by one or several frames, then the training target \cite{hawthorne2017onsets} will be completely changed.

In this work, we propose a regression-based high-resolution piano transcription system that can achieve arbitrary resolution in the time domain for music transcription. In training, we propose regression-based targets to represent the time difference between the centre of a frame and its nearest onset or offset times. The physical explanation is that each frame is assigned a target of \textit{how far} it is from its nearest onset or offset. Therefore, the information of precise onset or offset times have remained. At inference, we propose an analytical algorithm to calculate the precise onset or offset times with arbitrary time resolution. In evaluation, we investigate tolerances ranging from 2 ms to 100 ms for onset evaluation compared to the fixed tolerance of 50 ms \cite{hawthorne2017onsets}. We show that our proposed high-resolution piano transcription system achieves state-of-the-art results on the MAESTRO dataset \cite{hawthorne2018enabling}. We also show that our proposed regression-based targets are robust to the misaligned onsets and offsets. In addition, we develop a sustain pedal transcription system with our proposed method on the MAESTRO dataset, which has not been evaluated in previous works.

This paper is organized as follows. Section \ref{section:baseline} introduces neural network based piano transcription systems. Section \ref{section:proposed} introduces our proposed high-resolution piano transcription system. Section \ref{section:experiments} shows experimental results. Section \ref{section:conclusion} concludes this work. 

\section{Neural network-based piano transcription systems}\label{section:baseline}

\subsection{Frame-wise Transcription Systems}
Neural networks have been applied to tackle the piano transcription problem in previous works \cite{bock2012polyphonic, sigtia2016endtoend, kelz2016potential, hawthorne2017onsets}. First, audio recordings are transformed into log mel spectrograms as input features. We denote a log mel spectrogram as $ X \in \mathbb{R}^{T \times F} $, where $ T $ is the number of frames, and $ F $ is the number of mel frequency bins. Then, neural networks, such as fully connected neural networks, CNNs, or RNNs are applied on the log mel spectrograms to predict the frame-wise presence probabilities of piano notes. Usually, a frame-wise roll $ I_{\text{fr}} \in \{0, 1\}^{T \times K} $ is used as a target for training \cite{sigtia2016endtoend}, where $ K $ is the number of pitch classes, and is equivalent to $ 88 $ for piano transcription. The elements of $ I_{\text{fr}} $ have values of either 1 or 0, indicating the presence or absence of piano notes. Neural network-based methods \cite{sigtia2016endtoend} use a function $ f $ to map the log mel spectrogram of an audio clip to the frame-wise roll $ I_{\text{fr}} $. We denote the neural network output as $ P_{\text{fr}} = f(X) $, which has a shape of $ T \times K $. The function $ f $ is modeled by a neural network with a set of learnable parameters. The following loss function is used to train the neural network \cite{sigtia2016endtoend}:
\begin{equation} \label{eq:frames_loss}
l_{\text{fr}} = \sum_{t=1}^{T} \sum_{k=1}^{K} l_{\text{bce}}(I_{\text{fr}}(t, k), P_{\text{fr}}(t, k)),
\end{equation}
\noindent where $ l_{\text{fr}} $ is the frame-wise loss, and $ l_{\text{bce}}(\cdot, \cdot) $ is a binary cross-entropy function defined as:
\begin{equation} \label{eq:bce}
l_{\text{bce}}(y, p) = - y \text{ln} p - (1 - y) \text{ln} (1 - p).
\end{equation}
\noindent The target $ y \in \{0, 1\} $ is a binarized value and $ p \in [0, 1] $ is the predicted probability. A frame-wise piano transcription system is trained to predict the presence probability of notes in each frame \cite{sigtia2016endtoend}.

At inference, we first calculate the log mel spectrogram of an audio recording. Then, the log mel spectrogram is input to the trained neural network to calculate $ f(X) $. Finally, those predictions are post-processed to piano note events \cite{sigtia2016endtoend}.

\subsection{Onsets and Frames Transcription System}
For a piano, an onset is a hammer-string impact, which is equivalent to the beginning of a piano note event. An offset is a deactivated hammer-string impact \cite{goebl2001melody, goebl2003role}, for both MAESTRO dataset \cite{hawthorne2018enabling} and real-world recordings. One problem of the frame-wise transcription systems is that the transcribed frames need to be elaborately post-processed to piano notes \cite{sigtia2016endtoend}. In addition, those frame-wise piano transcription systems do not predict onsets explicitly, while the onsets carry rich information of piano notes. To address this problem, onsets and offsets dual objective system \cite{hawthorne2017onsets} was proposed to predict onsets and frames jointly. The onset and frame predictions are modeled by individual acoustic models containing several convolutional layers and long short term memory (LSTM) layers. The predicted onsets are used as conditional information to predict frame-wise outputs. We denote the predicted onset and frame outputs as $ P_\text{on} $ and $ P_\text{fr} $ respectively, where $ P_\text{on} $ and $ P_\text{fr} $ have shapes of $ T \times K $. We denote the onset and frame targets as $ I_\text{on} $ and $ I_\text{fr} $ respectively, where $ I_\text{on} $ and $ I_\text{fr} $ also have shapes of $ T \times K $. In \cite{hawthorne2017onsets}, a joint frame and onset loss function was used to train the onsets and frames system:
\begin{equation} \label{eq:onsets_frames_loss}
l_{\text{note}} = l_{\text{on}} + l_{\text{fr}},
\end{equation}
\noindent where $ l_{\text{fr}} $ is the frame-wise loss defined in (\ref{eq:frames_loss}), and $ l_{\text{on}} $ is the onset loss defined as:
\begin{equation} \label{eq:onsets_loss}
l_{\text{on}} = \sum_{t=1}^{T} \sum_{k=1}^{K} l_{\text{bce}}(I_{\text{on}}(t, k), P_{\text{on}}(t, k)).
\end{equation}
One advantage of the onsets and frames system is that the onset predictions can be used as extra information to predict frame-wise outputs. The onsets and frames system has become a benchmark system for piano transcription.

\section{high-resolution piano transcription system}\label{section:proposed}

\begin{figure}[t]
  \centering
  \centerline{\includegraphics[width=\columnwidth]{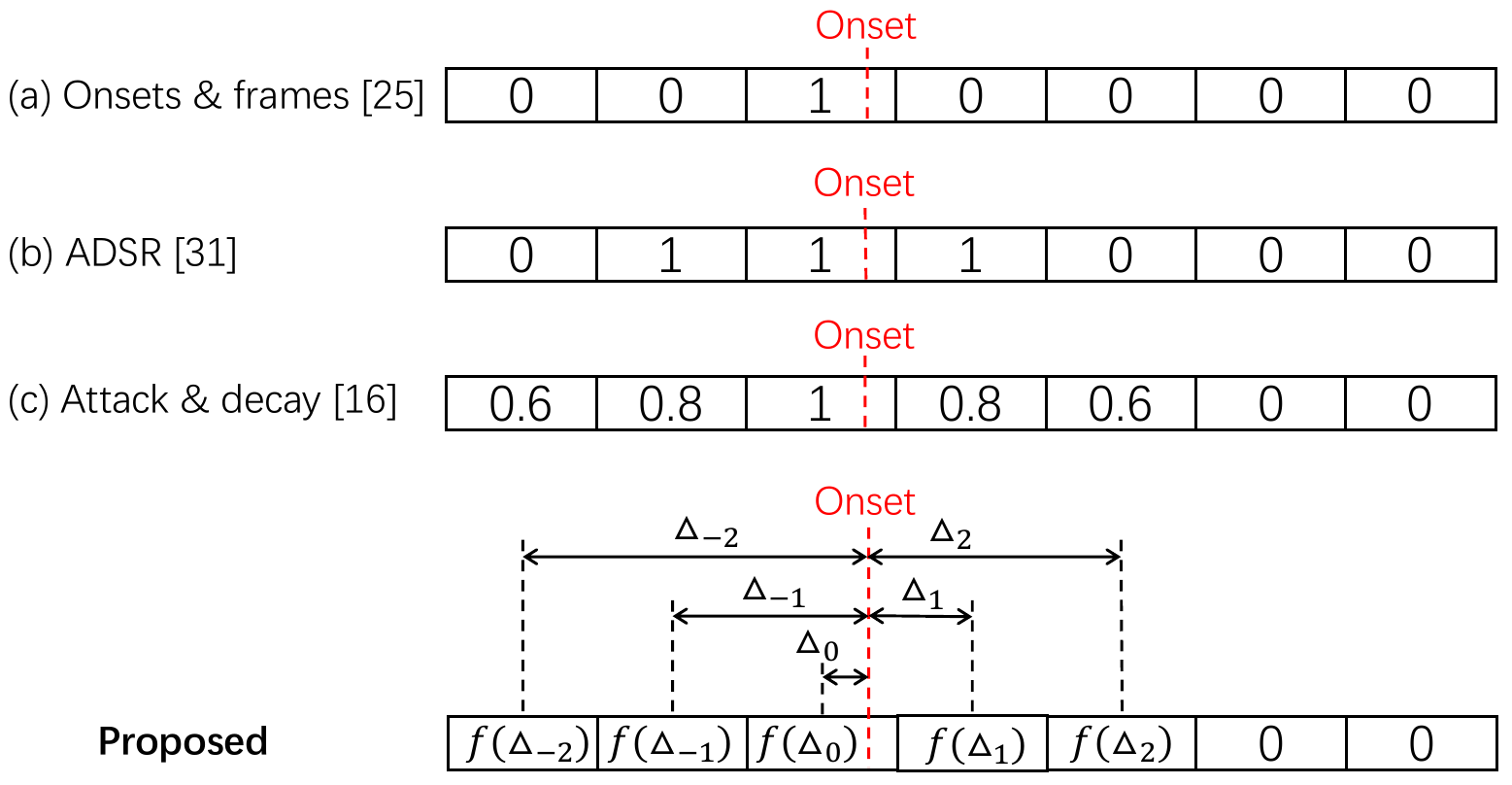}}
  \caption{Training targets of previous and our proposed piano transcription systems.}
  \label{fig:onset_to_target}
\end{figure}

Previous piano transcription systems introduced in Section \ref{section:baseline} have several limitations. The methods in Section \ref{section:baseline} predicts the presence or absence of onsets and frames in frames. Therefore, the transcription resolutions of those methods are limited to the hop size between adjacent frames. For example, the system \cite{hawthorne2017onsets} applies a hop size of 32 ms, so the transcription resolution in the time domain is limited to 32 ms. Second, for each piano note, previous systems \cite{hawthorne2017onsets} only label one or several frames of an onset or offset as 1, with other frames labelled as 0. The first row of Fig. \ref{fig:onset_to_target} shows the onset targets used in \cite{hawthorne2017onsets}. The dashed vertical line shows the precise onset time of a note. The onsets and frames system \cite{hawthorne2017onsets} only assign one positive value to several consecutive frames indicating the onset of a piano note. However, this can be imprecise because it is unclear how many frames an attack will last. The onsets and frames system \cite{hawthorne2017onsets} labelled two consecutive frames as onsets empirically, which may not be optimal. In \cite{kelz2019deep}, ADSR states are used to model the onsets and offsets, and several neighbouring frames of an onset are labelled as 1, as shown in the second row of Fig. \ref{fig:onset_to_target}. The offset time of notes will not be changed by reverberation, but the waveform of audio recordings can be blurred, which will lead to the difficulty of offset detection.

In addition, the targets shown in the first row of Fig. \ref{fig:onset_to_target} are sensitive to the misalignment of onset or offset labels. To explain, shifting the onset by one or several frames will lead to a completely different target. To mitigate this problem, Cheng et al. \cite{cheng2016attack} proposed attack and decay targets to model the onset of piano notes shown in the third row of Fig. \ref{fig:onset_to_target}. Instead of only labeling the frames containing onsets as 1, the neighbouring frames of onsets are labelled with continuous values. Similar ideas have been proposed to tackle the pitch estimation \cite{bittner2017deep} and music structure analysis \cite{ullrich2014boundary} problems using smoothing filters for boundaries prediction.

Another problem of previous onset target representations \cite{hawthorne2017onsets, kelz2019deep, cheng2016attack} is that they do not reflect the precise onset or offset times of notes. The precise onset and offset times information is lost when quantizing onset and offset times into frames. We explain this in the first to the third rows of Fig. \ref{fig:onset_to_target}. The onset targets are unchanged when the precise onset times (dashed lines) are shifted within a frame. To achieve high-resolution piano transcription, the onset targets should be sensitive to the onset shifts in milliseconds. Furthermore, when the precise onset time is on the boundary between two frames, the target can be confusing. In addition, efforts to increase transcription resolution in the time domain by simply reducing frame hop size will take more computation cost, and the problem of limited transcription resolution still exists.

\begin{figure}[t]
  \centering
  \centerline{\includegraphics[width=\columnwidth]{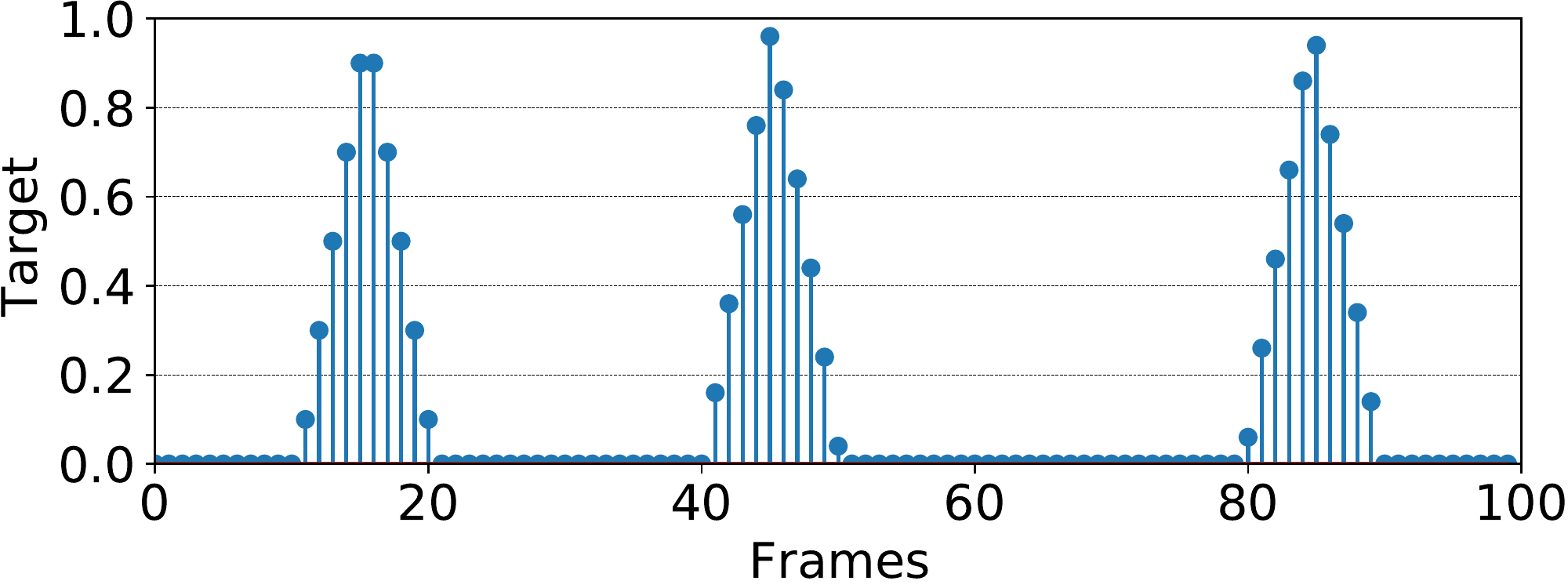}}
  \caption{High-resolution training targets of three notes with a same pitch.}
  \label{fig:regression_target}
\end{figure}

\subsection{Regress Onset and Offset Times}
We propose a high-resolution piano transcription system by predicting the continuous onset and offset times of piano notes instead of classifying the presence probabilities of onsets and offsets in each frame. This idea is inspired by the \textit{You Look Only Once} (YOLO) \cite{redmon2016you} object detection method from computer vision. In YOLO, an image is split into grids. Then, each grid predicts a distance between the coordinate of the grid and the coordinate of an object. The distance to be predicted is a continuous value. Different from previous works \cite{hawthorne2017onsets, kelz2019deep, cheng2016attack, bittner2017deep, ullrich2014boundary}, our proposed targets have a physical explanation that each frame is assigned a target of \textit{how far} it is from its nearest onset or offset. The targets are calculated by the time distance between the centre of a frame and the precise onset time of a note. The bottom row of Fig. \ref{fig:onset_to_target} shows the targets of our proposed high-resolution piano transcription system. We denote the frame hop size time as $ \Delta $, and the time difference between the centre of a frame and its nearest onset time as $ \Delta_{i} $, where $ i $ is the index of a frame. Negative $ i $ and positive $ i $ indicate previous and future frame indexes of an onset. Different from the targets of previous works \cite{hawthorne2017onsets, kelz2019deep, cheng2016attack} shown in the first to the third rows of Fig. \ref{fig:onset_to_target}, our proposed time difference $ \Delta_{i} $ contains precise onset and offset times information with arbitrary resolution. In training, we encode the time difference $ \Delta_{i} $ to targets $ g(\Delta_{i}) $ by a function $ g $:
\begin{equation} \label{eq:function_delta}
\left\{\begin{matrix}
g(\Delta_{i}) = 1 - \frac{|\Delta_{i}|}{J\Delta}, |i| \leq J \\ 
g(\Delta_{i}) = 0, |i| > J,
\end{matrix}\right.
\end{equation}
\noindent where $ J $ is a hyper-parameter controlling the sharpness of the targets. Larger $ J $ indicates ``smoother'' target, and smaller $ J $ indicates ``sharper'' target. Fig. \ref{fig:regression_target} 
shows the visualization of onset targets of a pitch with $ J = 5 $. There are three piano notes in Fig. \ref{fig:regression_target}. Different from the attack and decay targets \cite{cheng2016attack} shown in the third row of Fig. \ref{fig:onset_to_target}, the targets $ g(\Delta_{i}) $ in Fig. \ref{fig:regression_target} contain precise onset times information of piano notes. For example, Fig. \ref{fig:regression_target} shows that the onset of the first note is on the boundary of two adjacent frames, and the onsets of the second and third notes appear in different times. In training, both onset and offset regression targets are matrices with shapes of $ T \times K $. We denote onset and offset regression targets as $ G_{\text{on}} $ and $ G_{\text{off}} $ respectively, to distinguish them from the binarized targets $ I_{\text{on}} $ and $ I_{\text{off}} $ in Section \ref{section:baseline}. We denote the predicted onset and offset regression values as $ R_{\text{on}} $ and $ R_{\text{off}} $ respectively, to distinguish them from $ P_{\text{on}} $ and $ P_{\text{off}} $ in Section \ref{section:baseline}. Both of regression based outputs $ R_{\text{on}} $ and $ R_{\text{off}} $ have values between $ 0 $ and $ 1 $. We define the onset regression loss $ l_{\text{on}} $ and offset regression loss $ l_{\text{off}} $ as:
\begin{equation} \label{eq:regress_onset_loss}
l_{\text{on}} = \sum_{t=1}^{T} \sum_{k=1}^{K} l_{\text{bce}}(G_{\text{on}}(t, k), R_{\text{on}}(t, k)),
\end{equation}
\begin{equation} \label{eq:regress_offset_loss}
l_{\text{off}} = \sum_{t=1}^{T} \sum_{k=1}^{K} l_{\text{bce}}(G_{\text{off}}(t, k), R_{\text{off}}(t, k)).
\end{equation}
\noindent Equation (\ref{eq:regress_onset_loss}) and (\ref{eq:regress_offset_loss}) use regression-based targets instead of classification-based targets in (\ref{eq:onsets_loss}). To be consistent with the binary cross-entropy loss used in (\ref{eq:frames_loss}) and (\ref{eq:onsets_loss}), we use binary cross-entropy in (\ref{eq:regress_onset_loss}) and (\ref{eq:regress_offset_loss}). Equation (\ref{eq:regress_onset_loss}), and (\ref{eq:regress_offset_loss}) are minimized when $ R_{\text{on}} $ equals $ G_{\text{on}} $ and $ R_{\text{off}} $ equals $ G_{\text{off}} $. 

\begin{figure}[t]
  \centering
  \centerline{\includegraphics[width=0.8\columnwidth]{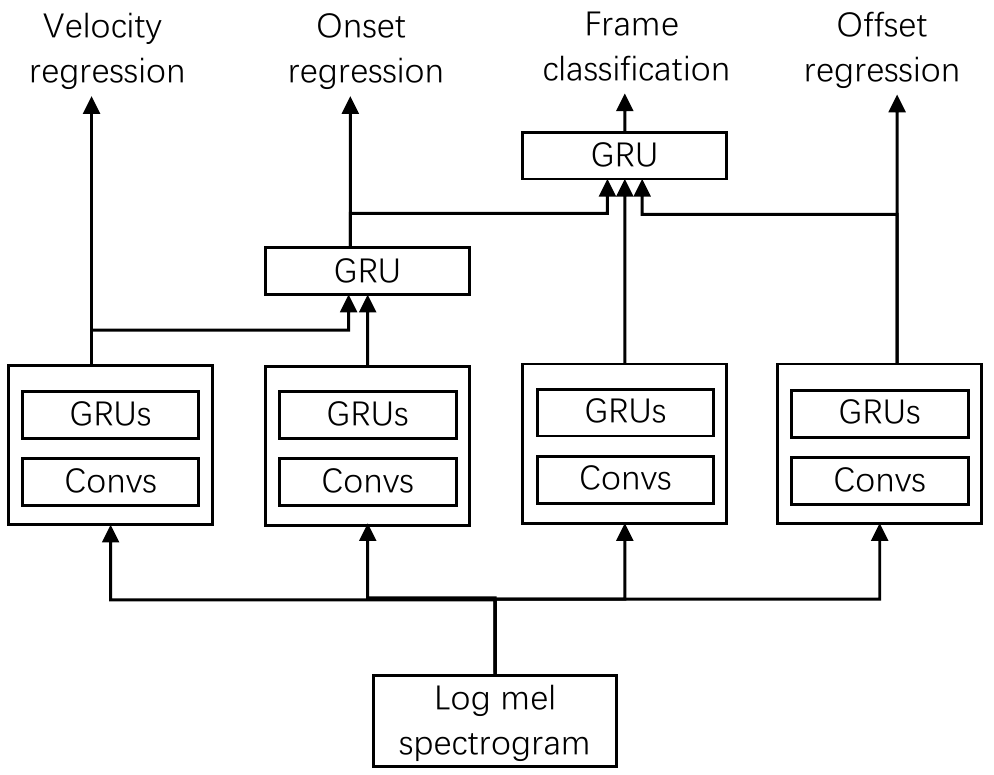}}
  \caption{High-resolution piano transcription system by regressing velocities, onsets, offsets and frames.}
  \label{fig:framework}
\end{figure}

\subsection{Velocity Estimation}
Velocities of piano notes are correlated with the loudness of piano notes being played. Dannenberg and Goebl analyzed the correlation between velocity and loudness in \cite{dannenberg2006interpretation} and \cite{goebl2003role}. In this work, we estimate the velocity of notes via estimating the MIDI velocities \cite{hawthorne2018enabling}. We build a velocity estimation submodule to estimate the velocities of transcribed notes. MIDI files represent velocities of notes using integers ranging from 0 to 127. Larger integers indicate loud notes and smaller integers indicate quiet notes. To begin with, we normalize the dynamic range of velocities from $ [0, 127] $ to $ [0, 1] $. We denote the ground truth and predicted velocities as $ I_{\text{vel}} $ and $ P_{\text{vel}} $ respectively, where $ I_{\text{vel}} $ and $ P_{\text{vel}} $ have shapes of $ T \times K $. Then, we define the velocity loss as:
\begin{equation} \label{eq:velocity_loss}
l_{\text{vel}} = \sum_{t=1}^{T} \sum_{k=1}^{K} I_{\text{on}}(t, k) \cdot l_{\text{bce}}(I_{\text{vel}}(t, k), P_{\text{vel}}(t, k)).
\end{equation}
\noindent Equation (\ref{eq:velocity_loss}) shows that the ground truth onsets $ I_{\text{on}}(t, k) $ are used to modulate the velocity prediction. That is, we only predict velocities for onsets. One motivation is that the onset of piano notes carry rich information of their velocities, while the decay of piano notes carry less information of velocities than onsets. Similar to (\ref{eq:regress_onset_loss}) and (\ref{eq:regress_offset_loss}), binary cross-entropy is used to optimize (\ref{eq:velocity_loss}). The loss $ l_{\text{vel}} $ is minimized when $ P_{\text{vel}}(t, k) $ equals $ I_{\text{vel}}(t, k) $. At inference, we only predict velocities where onsets are detected. Finally, the predicted velocities are scaled from $ [0, 1] $ back to integers of $ [0, 127] $.

\subsection{Entire System}\label{section:entire_system}
\begin{figure}[t]
  \centering
  \centerline{\includegraphics[width=0.8\columnwidth]{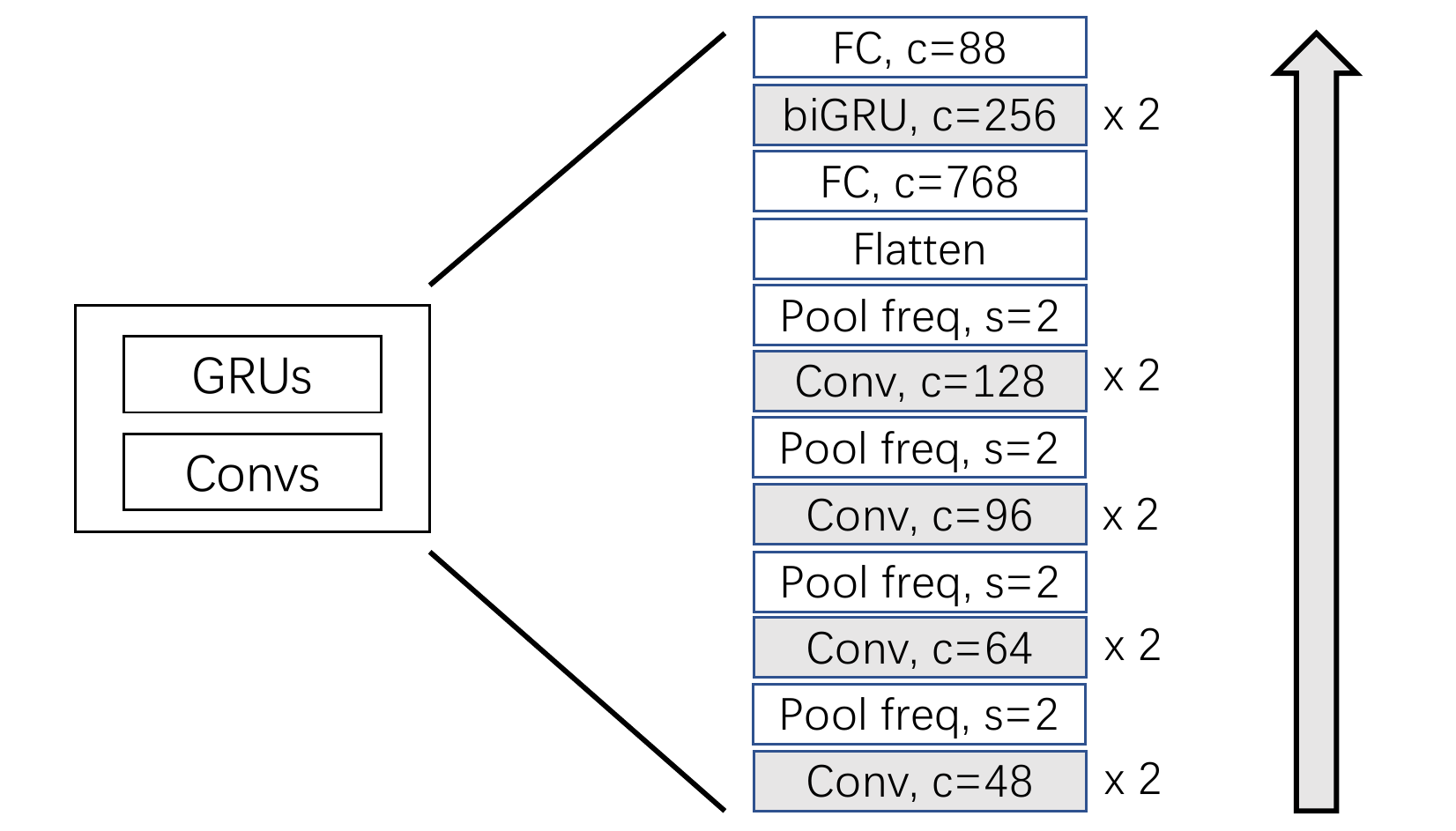}}
  \caption{Acoustic model. Pooling is only applied along the frequency axis. Letter ``c'' and ``s'' indicate the number of channels and the downsampling rate.}
  \label{fig:conv_gru_block}
\end{figure}

Fig. \ref{fig:framework} shows the framework of our proposed high-resolution piano transcription system. First, an audio clip is transformed into a log mel spectrogram with a shape of $ T \times F $ as input feature \cite{hawthorne2017onsets}, where $ F $ is the number of mel frequency bins. There are four submodules in Fig. \ref{fig:framework} from left to right: a velocity regression submodule, an onset regression submodule, a frame-wise classification submodule, and an offset regression submodule. Each submodule is modeled by an acoustic model \cite{hawthorne2017onsets}. In our system, we model each acoustic model with several convolutional layers followed by bidirectional gated recurrent units (biGRU) layers. The convolutional layers are used to extract high-level information from the log mel spectrogram, and the biGRU layers are used to summarize long-time information of the log mel spectrogram. Then, a time-distributed fully connected layer is applied after the biGRU layer to predict regression results. The outputs of all acoustic models have dimensions of $ T \times K $. Fig. \ref{fig:conv_gru_block} shows the neural network layers of the acoustic models used in Fig. \ref{fig:framework}. We will describe the detailed configuration of the acoustic models in Section \ref{section:model_architecture}.

Fig. \ref{fig:framework} shows that the predicted velocities are used as conditional information to predict onsets. One motivation is that the velocity of a piano note can be helpful to detect its corresponding onset. For example, if the velocity of a note is low, then the system will attend more to the frame to detect the onset. This simulates human beings who will listen carefully to the notes with low velocity. We concatenate the outputs of the velocity regression submodule and the onset regression submodule along the frequency dimension and use this concatenation as an input to a biGRU layer to calculate the final onset predictions. Similarly, we concatenate the outputs of the onset regression and offset regression submodules and use this concatenation as the input to a biGRU layer to calculate the frame-wise predictions. The total loss function to train our proposed piano transcription system consists of four parts:
\begin{equation} \label{eq:total_loss}
l_{\text{note}} = l_{\text{fr}} + l_{\text{on}} + l_{\text{off}} + l_{\text{vel}},
\end{equation}
\noindent where $ l_{\text{fr}} $, $ l_{\text{on}} $, $ l_{\text{off}} $ and $ l_{\text{velocity}} $ are described in (\ref{eq:frames_loss}), (\ref{eq:regress_onset_loss}), (\ref{eq:regress_offset_loss}) and (\ref{eq:velocity_loss}) respectively. We simply weight all those losses equally, which works well in our experiment.

\subsection{Inference}\label{section:inference}
At inference, we input the log mel spectrogram of an audio recording into the trained piano transcription system to calculate the frame-wise prediction, onset regression, offset regression, and velocity regression outputs. Then, we propose an algorithm to process those outputs to high-resolution note events, where each note event can be represented by a quadruple of <piano note, onset time, offset time, velocity>.

\begin{figure}[t]
  \centering
  \centerline{\includegraphics[width=0.8\columnwidth]{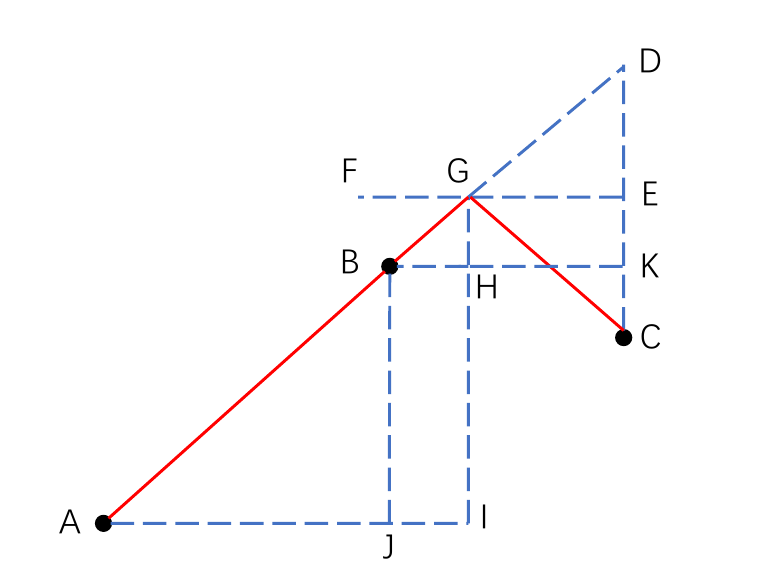}}
  \caption{Demonstration of calculating the precise onset or offset time of a note. The points $ A $, $ B $ and $ C $ are the predictions of three frames. The point $ G $ is the calculated precise onset or offset time.}
  \label{fig:inference}
\end{figure}

Fig. \ref{fig:inference} shows the strategy of calculating precise onset or offset times of piano notes. The horizontal coordinate of $ A $, $ B $, $ C $ are the centre times of three adjacent frames, where $ B $ is the frame with a local maximum onset prediction value. Previous works \cite{hawthorne2017onsets} regard the time of $ B $ as the onset time. In this case, the transcription resolution is limited to the hop size between adjacent frames. Ideally, the precise onset time should be $ G $. We propose to calculate the precise onset time of $ G $ as follows. First, we detect $ A $, $ B $, $ C $ where $ B $ is a local maximum frame. If the vertical value of $ B $ is larger than an onset threshold, then we say there exists an onset near the frame $ B $. Next, we analytically calculate the precise time of $ G $ where $ G $ satisfies $ AG $ and $ CG $ are symmetric along the vertical line $ GI $.

Without loss of generality, we assume the output value of $ C $ is larger than $ A $. We denote the coordinate of $A$, $B$ and $C$ as $ (x_{A}, y_{A}) $, $ (x_{B}, y_{B}) $ and $ (x_{C}, y_{C}) $ respectively. We show that the time difference between $ B $ and $ G $ is:
\begin{equation} \label{eq:precise_1}
BH = \frac{x_{B} - x_{A}}{2} \frac{y_{C} - y_{A}}{y_{B} - y_{A}}.
\end{equation}

\noindent \textit{Proof.} Extend $ AB $ to $ D $ where $ CD $ is a vertical line. Take the median point of $ CD $ as $E$. Draw a horizontal line $ EF $ cross $ AD $ at $ G $. Then, we calculate $ BH $. We have $ \bigtriangleup BGH \sim \bigtriangleup ABJ $, so $ BH = \frac{AJ \cdot GH}{BJ} $. We know that $ AJ = x_{B} - x_{A} $, $ BJ = y_{B} - y_{A} $ and $ GH = DK - DE $, and we know $ AJ = BK $, so $ \bigtriangleup ABJ \cong \bigtriangleup BDK $, so $ DK = y_{B} - y_{A} $. Then, we have $ DE = \frac{CD}{2} = \frac{DK + CK}{2} = y_{B} - \frac{y_{A} + y_{C}}{2} $, so $ GH = \frac{y_{C} - y_{A}}{2} $, so $ BH = \frac{x_{B} - x_{A}}{2} \frac{y_{C} - y_{A}}{y_{B} - y_{A}} $. $ \square $

\noindent In another case, if the output value of $ A $ is larger than $ C $, then:
 \begin{equation} \label{eq:precise_2}
BH = \frac{x_{C} - x_{B}}{2} \frac{y_{A} - y_{C}}{y_{B} - y_{C}}.
\end{equation}
 
 \noindent By this means, we can calculate the precise onset or offset times of piano notes. We describe onset and offset times detection in Algorithm \ref{alg:note_transcription}. A frame is detected to contain an onset if they are over an onset threshold $ \theta_{\text{on}} $ and is a local maximum. Then, the precise onset time is calculated by (\ref{eq:precise_1}) or (\ref{eq:precise_2}). The velocity of an onset is obtained by scaling the predicted velocity from $ [0, 1] $ to a range of $ [0, 127] $. For each detected onset, an offset is detected if the offset regression output is over an offset threshold $ \theta_{\text{off}} $ or any frame prediction outputs are lower than a frame threshold $ \theta_{\text{fr}} $. When consecutive onsets of the same pitch are detected, their previous onsets are truncated by adding offsets. Our proposed method can not detect consecutive notes shorter than 4 frames (40 ms) because of the local maximum detection algorithm. Still, repeating a note shorter than 40 ms rarely happens in real piano performance. Finally, all onsets and offsets are paired to constitute piano notes.

\begin{algorithm}[t]
	\caption{Onset and offset times detection.}\label{alg:note_transcription}
	\begin{algorithmic}[1]
		\State Inputs: $ R_{\text{on}}(t, k) $, $ R_{\text{off}}(t, k) $, $ P_{\text{fr}}(t) $, $ P_{\text{vel}}(t, k) $, $ \theta_{\text{on}} $, $ \theta_{\text{off}} $ and $ \theta_{\text{fr}} $.
	    \State Outputs: Detected onset and offset times.
	    \For{$k = 1, ..., K$}
	    \For{$t = 1, ..., T$}
	        \\
	        \ \ \ \ \ \ \ \ \textit{\# Detect note onset.}
	        \If{$ R_{\text{on}}(t, k) > \theta_{\text{on}} $ and $ R_{\text{on}}(t, k) $ is local maximum}
	            \State Note onset of pitch $ k $ is detected. The precise onset time is refined by (\ref{eq:precise_1}) or (\ref{eq:precise_2}).
	            \State Calculate the velocity of the note by $ P_{\text{vel}}(t, k) \times 128 $.
	        \EndIf
	        \\
	        \ \ \ \ \ \ \ \ \textit{\# Detect note offset.}
	        \If{($R_{\text{off}}(t, k) > \theta_{\text{off}}$ and $ R_{\text{off}}(t, k) $ is local maximum) or $P_{\text{fr}}(t, k) < \theta_{\text{fr}}$}
	        \State Note offset is $ k $ detected. The precise offset time is refined by (\ref{eq:precise_1}) or (\ref{eq:precise_2}).
	        \EndIf
	    \EndFor
	    \EndFor
	        
	\end{algorithmic}
\end{algorithm}

\subsection{Sustain Pedal Transcription}\label{section:pedal}
In this section, we will show our proposed regression-based transcription system can be applied to sustain pedal detection. The sustain pedal transcription system is similar to the piano note transcription system. The only difference is that the sustain pedal transcription system has output shapes of $ T \times 1 $ instead of $ T \times 88 $, where the outputs indicate the presence probability of sustain pedals. Sustain pedals are important for piano performance. When pressed, the sustain pedal sustains all damped strings on a piano by moving all dampers away from the strings and allows strings to vibrate freely. All notes being played will continue to sound until the pedal is released. However, many previous piano transcription systems \cite{hawthorne2017onsets, cheng2016attack, kelz2019deep} did not incorporate sustain pedal transcription. On the other hand, some sustain pedal transcription systems \cite{liang2019piano} do not include piano notes transcription. In \cite{liang2019piano}, a convolutional neural network was used to detect piano pedals in each frame. Still, there is a lack of benchmark sustain pedal transcription systems on the MAESTRO dataset \cite{hawthorne2018enabling}.

In this section, we propose a sustain pedal transcription system with our proposed high-resolution transcription system. The sustain pedal transcription system is separate from the note transcription system and is trained separately. Training the note and sustain pedal systems separately leads to better transcription performance and also has the advantage of reducing memory usage. After training the note transcription and sustain pedal transcription systems, we combine them into a unified model for release. In the MIDI format, sustain pedals are represented with integer values ranging from 0 to 128. MIDI values larger than 64 are regarded as ``on'' and MIDI values smaller than 64 are regarded as ``off''.

To simplify the sustain pedal transcription problem, we only classify the ``on'' and ``off'' states of sustain pedals and do not consider advanced sustain pedal techniques such as half pedals. We denote the pedal onset regression target, offset regression target, and frame-wise target as $ G_{\text{ped\_on}} \in [0, 1]^{T} $, $ G_{\text{ped\_off}} \in [0, 1]^{T} $, and $ I_{\text{ped\_fr}} \in \{0, 1\}^{T} $, respectively. The onset and offset regression targets $ G_{\text{ped\_on}} $ and $ G_{\text{ped\_off}} $ are obtained by (\ref{eq:function_delta}), and have continuous values between 0 and 1. The frame-wise targets have binarized values between 0 and 1. We apply acoustic models described in Section \ref{section:entire_system} to predict the onset regression, offset regression and frame-wise outputs of pedals. We denote the predicted onsets, offsets, and frame-wise values as $ R_{\text{ped\_on}}(t) $, $ R_{\text{ped\_off}}(t) $ and $ P_{\text{ped\_fr}}(t) $, respectively. We use the following loss function to train the sustain pedal transcription system:
\begin{equation} \label{eq:pedal_onset_loss}
l_{\text{ped\_on}} = \sum_{t=1}^{T}l_{\text{bce}}(G_{\text{ped\_on}}(t), R_{\text{ped\_on}}(t)),
\end{equation}
\begin{equation} \label{eq:pedal_offset_loss}
l_{\text{ped\_off}} = \sum_{t=1}^{T}l_{\text{bce}}(G_{\text{ped\_off}}(t), R_{\text{ped\_off}}(t)),
\end{equation}
\begin{equation} \label{eq:pedal_frame_loss}
l_{\text{ped\_fr}} = \sum_{t=1}^{T}l_{\text{bce}}(I_{\text{ped\_fr}}(t), P_{\text{ped\_fr}}(t)).
\end{equation}
\noindent Then, the total loss function is calculated by:
\begin{equation} \label{eq:pedal_loss}
l_{\text{ped}} = l_{\text{ped\_fr}} + l_{\text{ped\_on}} + l_{\text{ped\_off}}.
\end{equation}
\noindent At inference, we propose Algorithm \ref{alg:pedal_transcription} to process the sustain pedal prediction outputs into sustain pedal events. Sustain pedal onsets detection is different from notes onset detection. The press of a sustain pedal is usually before the press of piano notes so $ R_{\text{ped\_on}}(t) $ can be difficult to detect. We design a sustain pedal detection system so that a pedal onset is detected when the frame-wise prediction $ R_{\text{ped\_fr}}(t) $ is over a threshold $ \theta_{\text{ped\_on}} $. A pedal offset is detected if the pedal offset prediction $ R_{\text{ped\_off}}(t) $ is higher than an offset threshold $ \theta_{\text{ped\_off}} $ or the frame-wise prediction $ P_{\text{ped\_fr}}(t) $ is lower than a threshold $ \theta_{\text{ped\_fr}} $. 

\begin{algorithm}[t]
	\caption{Sustain pedal onset and offset times detection.}\label{alg:pedal_transcription}
	\begin{algorithmic}[1]
		\State Inputs: $ R_{\text{off}}(t) $, $ R_{\text{fr}}(t) $, $ \theta_{\text{ped\_off}} $, $ \theta_{\text{ped\_fr}} $.
	    \State Outputs: Detected pedal onset and offset times.
	    \For{$t = 1, ..., T$}
	        \\
	        \ \ \ \ \textit{\# Detect pedal onset.}
	        \If{$ R_{\text{ped\_fr}}(t) > \theta_{\text{ped\_on}} $ and $ R_{\text{ped\_fr}}(t) > R_{\text{ped\_fr}}(t-1) $}
	            \State Pedal onset is detected.
	        \EndIf
	        \\
	        \ \ \ \ \textit{\# Detect pedal offset.}
	        \If{$R_{\text{ped\_off}}(t) > \theta_{\text{ped\_off}}$ or $P_{\text{ped\_fr}}(t) < \theta_{\text{ped\_fr}}$}
	        \State Pedal offset is detected. The precise offset time is refined by (\ref{eq:precise_1}) or (\ref{eq:precise_2}).
	        \EndIf
	    \EndFor
	        
	\end{algorithmic}
\end{algorithm}

\section{Experiments}\label{section:experiments}

\subsection{Dataset}
To compare with previous piano transcription systems, we use the MAESTRO dataset V2.0.0 \cite{hawthorne2018enabling}, a large-scale dataset containing paired audio recording and MIDI files to train and evaluate our proposed piano transcription system. The MAESTRO dataset contains piano recordings from the International Piano-e-Competition. Pianists performed on Yamaha Disklaviers concert-quality acoustic grand pianos integrated with high-precision MIDI capture and playback system. The MAESTRO dataset contains over 200 hours of solo piano recordings. Those audio recordings and MIDI files are aligned with a time resolution of around 3 ms introduced by \cite{hawthorne2018enabling}. Each music recording contains meta-information, including the composer, title, and year of the performance. MAESTRO dataset consists of training, validation and testing subsets.

\subsection{Preprocessing}
\begin{figure}[t]
  \centering
  \centerline{\includegraphics[width=\columnwidth]{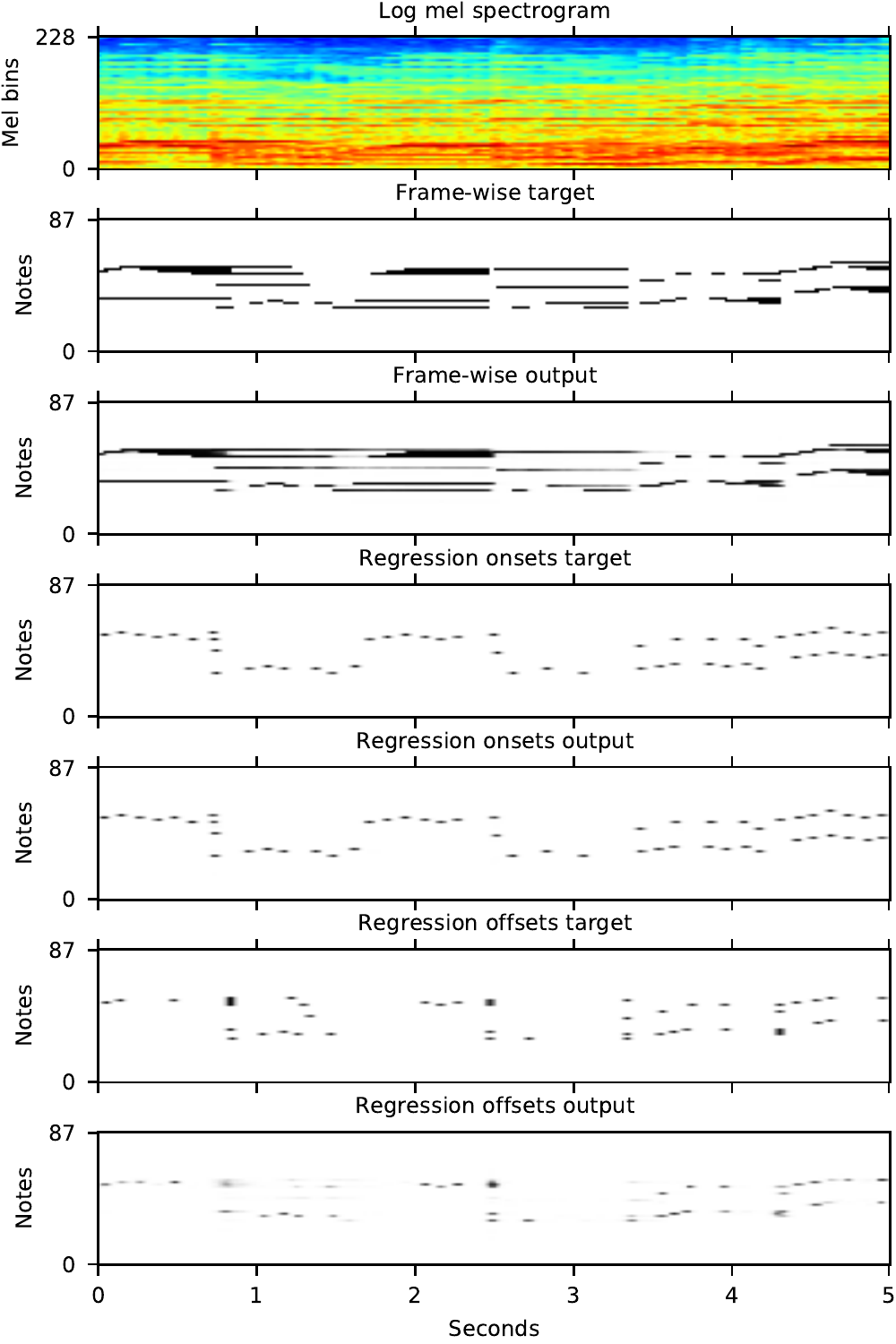}}
  \caption{From top to bottom: Log mel spectrogram of a 5-second audio clip; frame-wise targets; frame-wise outputs; onset regression targets; onset regression outputs; offset regression targets; offset regression outputs. The music segment is: Prelude and Fugue in D Major, WTC I, BWV 850, J. S. Bach, 2'35.}
  \label{fig:note_outputs}
\end{figure}

We use Python and PyTorch deep learning toolkit \cite{paszke2019pytorch} to develop our systems. All stereo audio recordings are converted into mono and are resampled to 16 kHz following \cite{hawthorne2018enabling}. The cutoff frequency of 16 kHz covers the frequency of the highest note $ \text{C}_{8} $ on a piano of 4186 Hz. We split audio recordings into 10-second clips. Then, a short-time Fourier transform with a Hann window size 2048 is used to extract the spectrogram. Mel banks with 229 banks and cutoff frequencies between 30 Hz and 8000 Hz are used to extract the log mel spectrogram \cite{hawthorne2018enabling}. We use a hop size of 10 ms between frames. For a 10-second audio clip, an input log mel spectrogram has a shape of $ 1001 \times 229 $, where the extra one frame comes from that audio clips are padded with half windows in both sides of a signal before feature extraction. Log mel spectrograms are extracted on the fly using the TorchLibrosa toolkit \cite{kong2019panns}. For general experiments, we set the hyper-parameter $ J = 5 $. That is, each onset or offset will affect the regression values of $ 2 \times J = 10 $ frames. We also investigate piano transcription systems with different $ J $ in our experiments. The first row of Fig. \ref{fig:note_outputs} shows an example of the log mel spectrogram of a 5-second audio clip. The second and third rows show the frame-wise targets and frame-wise predictions of the audio clip. The fourth and fifth rows show the regression onset targets and onset predictions of the audio clip. The sixth and seventh rows show the regression offset targets and offset predictions of the audio clip.

\begin{table*}
  \caption{Transcription results evaluated on the MAESTRO dataset}
  \vspace{6pt}
  \label{tab:total}
  \centering
  \resizebox{\textwidth}{!}{%
  \begin{tabular}{lcccccccccccc}
    \toprule
    & \multicolumn{3}{c}{\textbf{\textsc{Frame}}} & \multicolumn{3}{c}{\textbf{\textsc{Note}}} & \multicolumn{3}{c}{\textbf{\textsc{Note w/ offset}}} & \multicolumn{3}{c}{\textbf{\textsc{Note w/ offset \& vel.}}} \\
	\cmidrule(lr){2-4} \cmidrule(lr){5-7} \cmidrule(lr){8-10} \cmidrule(lr){11-13} 
     & P (\%) & R (\%) & F1 (\%) & P (\%) & R (\%) & F1 (\%) & P (\%) & R (\%) & F1 (\%) & P (\%) & R (\%) & F1 (\%) \\
    \midrule
    Onsets \& frames \cite{hawthorne2018enabling} & 93.10 & 85.76 & 89.19 & 97.42 & 92.37 & 94.80 & 81.84 & 77.66 & 79.67 & 78.11 & 74.13 & 76.04 \\
    Onsets \& frames [reproduced] & 86.63 & 90.89 & 88.63 & 99.52 & 89.23 & 93.92 & 80.43 & 72.27 & 75.99 & 79.51 & 71.48 & 75.14 \\
    Adversarial onsets \& frames \cite{kim2019adversarial} & 93.10 & 89.80 & \textbf{91.40} & 98.10 & 93.20 & 95.60 & 83.50 & 79.30 & 81.30 & 82.30 & 78.20 & 80.20 \\
    \midrule
    Regress cond on. & 86.94 & 90.15 & 88.42 & 98.43 & 94.84 & 96.57 & 80.00 & 77.08 & 78.50 & 78.64 & 75.79 & 77.17 \\
    Regress cond on. \& off. & 88.91 & 90.28 & 89.51 & 98.53 & 94.81 & 96.61 & 83.81 & 80.70 & 82.20 & 82.36 & 79.33 & 80.79 \\
    Regress cond on. \& off. \& vel. & 88.71 & 90.73 & 89.62 & 98.17 & 95.35 & \textbf{96.72} & 83.68 & 81.32 & \textbf{82.47} & 82.10 & 79.80 & \textbf{80.92} \\    
	\bottomrule
\end{tabular}}
\end{table*}

\begin{table*}
  \caption{Transcription results evaluated with different hyper-parameter $J$}
  \vspace{6pt}
  \label{tab:hyperparameter_J}
  \centering
  \resizebox{\textwidth}{!}{%
  \begin{tabular}{lcccccccccccc}
    \toprule
    & \multicolumn{3}{c}{\textbf{\textsc{Frame}}} & \multicolumn{3}{c}{\textbf{\textsc{Note}}} & \multicolumn{3}{c}{\textbf{\textsc{Note w/ offset}}} & \multicolumn{3}{c}{\textbf{\textsc{Note w/ offset \& vel.}}} \\
	\cmidrule(lr){2-4} \cmidrule(lr){5-7} \cmidrule(lr){8-10} \cmidrule(lr){11-13} 
     & P (\%) & R (\%) & F1 (\%) & P (\%) & R (\%) & F1 (\%) & P (\%) & R (\%) & F1 (\%) & P (\%) & R (\%) & F1 (\%) \\
    \midrule
    Hyper-parameter $J=2$ & 85.36 & 91.76 & 88.35 & 98.94 & 93.68 & 96.19 & 81.42 & 77.15 & 79.19 & 80.02 & 75.85 & 77.84 \\
    Hyper-parameter $J=5$ & 87.62 & 90.88 & \textbf{89.14} & 98.15 & 95.15 & \textbf{96.61} & 82.92 & 80.42 & \textbf{81.63} & 81.35 & 78.92 & \textbf{80.10} \\
    Hyper-parameter $J=10$ & 86.89 & 90.89 & 88.75 & 97.60 & 95.39 & 96.47 & 81.48 & 79.66 & 80.55 & 79.82 & 78.05 & 78.92 \\
    Hyper-parameter $J=20$ & 86.56 & 90.41 & 88.34 & 96.23 & 95.15 & 95.67 & 77.79 & 76.95 & 77.36 & 76.24 & 75.43 & 75.82 \\
	\bottomrule
\end{tabular}}
\end{table*}

\subsection{Model Architecture}\label{section:model_architecture}
After extracting the log mel spectrogram features, we apply a batch normalization \cite{ioffe2015batch} layer immediately on the individual frequency bins of the log mel spectrogram \cite{kong2019panns} to standardize the input. Then, acoustic models shown in Fig. \ref{fig:framework} modeled by CRNNs are applied to predict the velocity regression, onset regression, frame-wise classification, and offset regression outputs. All acoustic models have the same architecture, where each acoustic model consists of four convolutional blocks and two bidirectional biGRU layers. Each convolutional block consists of two convolutional layers with kernel sizes $ 3 \times 3 $. Batch normalization \cite{ioffe2015batch} and ReLU nonlinearity \cite{nair2010rectified} are applied after each linear convolutional operation to stabilize training and to increase the nonlinear representation ability of the system. The four convolutional blocks have output feature map numbers of $ 48 $, $ 64 $, $ 92 $ and $ 128 $ respectively. After each convolutional block, feature maps are averagely pooled by a factor of $ 2 $ along the frequency axis to reduce the feature map sizes. We do not apply pooling along the time axis to retain the transcription resolution in the time domain. 

After convolutional layers, feature maps are flattened along the frequency and channel axes and are input to a fully connected layer with 768 output units. Then, two biGRU layers with hidden sizes $ 256 $ are applied, followed by an additional fully connected layer with $ 88 $ sigmoid outputs. Dropout \cite{srivastava2014dropout} with rates of $ 0.2 $ and $ 0.5 $ are applied after convolutional blocks and fully connected layers to prevent the systems from overfitting. Fig. \ref{fig:framework} shows that the velocity and onset regression outputs are concatenated and are input to a biGRU layer. The biGRU layer contains $ 256 $ hidden units and is followed by a fully-connected layer with $ 88 $ sigmoid outputs to predict the final regression onsets. Similarly, the onset regression, offset regression, and frame-wise classification outputs are concatenated and are input to a biGRU layer. The biGRU contains $ 256 $ hidden outputs and is followed by a fully-connected layer with $ 88 $ sigmoid outputs to predict the final frame-wise output. The note transcription system consists of 20,218,778 trainable parameters. The training of the note transcription system applies a loss function described in (\ref{eq:total_loss}). The sustain pedal transcription submodule has the same acoustic model architecture as the note transcription submodule, except there is only one output instead of 88 outputs. The pedal onset regression, offset regression, and frame-wise classification are modeled by individual acoustic models. The training of the sustain pedal transcription system applies a loss function described in (\ref{eq:pedal_loss}).

We use a batch size $ 12 $, and an Adam \cite{kingma2014adam} optimizer with a learning rate of $ 0.0005 $ for training. The learning rate is reduced by a factor of $ 0.9 $ every $ 10 $ k iterations in training. Systems are trained for $200$ k iterations. The training takes four days on a single Tesla-V100-PCIE-32GB GPU card. At inference, we set onset, offset, frame-wise, and pedal thresholds to 0.3. All hyper-parameters are tuned on the validation set. The outputs are post-processed to MIDI events described in \ref{section:inference}.

\subsection{Evaluation}
We evaluate our proposed piano transcription system on the test set of the MAESTRO dataset. We compare our system with previous onsets and frames system \cite{hawthorne2017onsets} and the adversarial onsets and frames system \cite{kim2019adversarial}. The system \cite{hawthorne2017onsets} is an improvement to the onsets and frames system \cite{hawthorne2017onsets} and also used quantized targets for training. For a fair comparison with our proposed system with previous systems, we re-implemented the onsets and frames system trained with hard labels of 0 and 1. The results are shown in the third row of Table \ref{tab:total}. The numbers of our re-implemented system are slightly different from \cite{hawthorne2017onsets} due to different data augmentation strategies, different data pre-processing, data post-processing strategies, and deep learning toolkits. There are four types of evaluation metrics for piano transcription evaluation, including frame-wise evaluation, note evaluation with onset, note evaluation with both onset and offset, and note evaluation with onset, offset, and velocity. Following \cite{hawthorne2017onsets}, a tolerance of 50 ms is used for onset evaluation. A tolerance of 50 ms and an offset ratio of 0.2 are used for offset evaluation. A velocity tolerance of 0.1 is used for velocity evaluation, which indicates that estimated notes are considered correct if, after scaling and normalization velocities to a range of 0 to 1, they are within the velocity tolerance of a matched reference note. 

The first to the third rows of Table \ref{tab:total} show the results of the onsets and offsets system \cite{hawthorne2018enabling}, the reproduced onsets and offsets system, and the adversarial system \cite{kim2019adversarial} for a fair comparison. The fourth row shows that our proposed regression-based system only using onset as the condition improves the onsets and frames system \cite{hawthorne2018enabling} note F1 score from 94.80\% to 96.57\%. The fifth row shows that using both onset and offset as conditions achieves an onset F1 score of 96.61\%. The sixth row shows that using onset and offset to condition the frame prediction and using velocity to condition the onset prediction further improves the note F1 score to 96.72\%. \qk{We also evaluated the system performance of different runs and observed that the standard variance of frame F1 scores, onset F1 scores, onset F1 scores evaluated with offsets, and onset F1 scores evaluated with offsets and velocities are \textpm 0.04\%, 0.03\%, 0.08\%, and 0.10\%.} Our proposed high-resolution system improves the note F1 score evaluated with offsets from 79.67\% to 82.47\% and improves the note F1 score evaluated with both offset and velocity from 76.04\% to 80.92\%. The first row of Fig. \ref{fig:note_outputs} shows the log mel spectrogram of an audio clip. The second and third rows show the frame-wise target and frame-wise system output. The fourth and fifth rows show the regression onset target and regression onset output. The sixth and seventh row show the regression offset target and regression offset output. The onset and offset regression targets are calculated by (\ref{eq:function_delta}). Fig. \ref{fig:note_outputs} shows that our proposed system performs well on transcribing a 5-second audio clip. 

\begin{figure}[t]
  \centering
  \centerline{\includegraphics[width=\columnwidth]{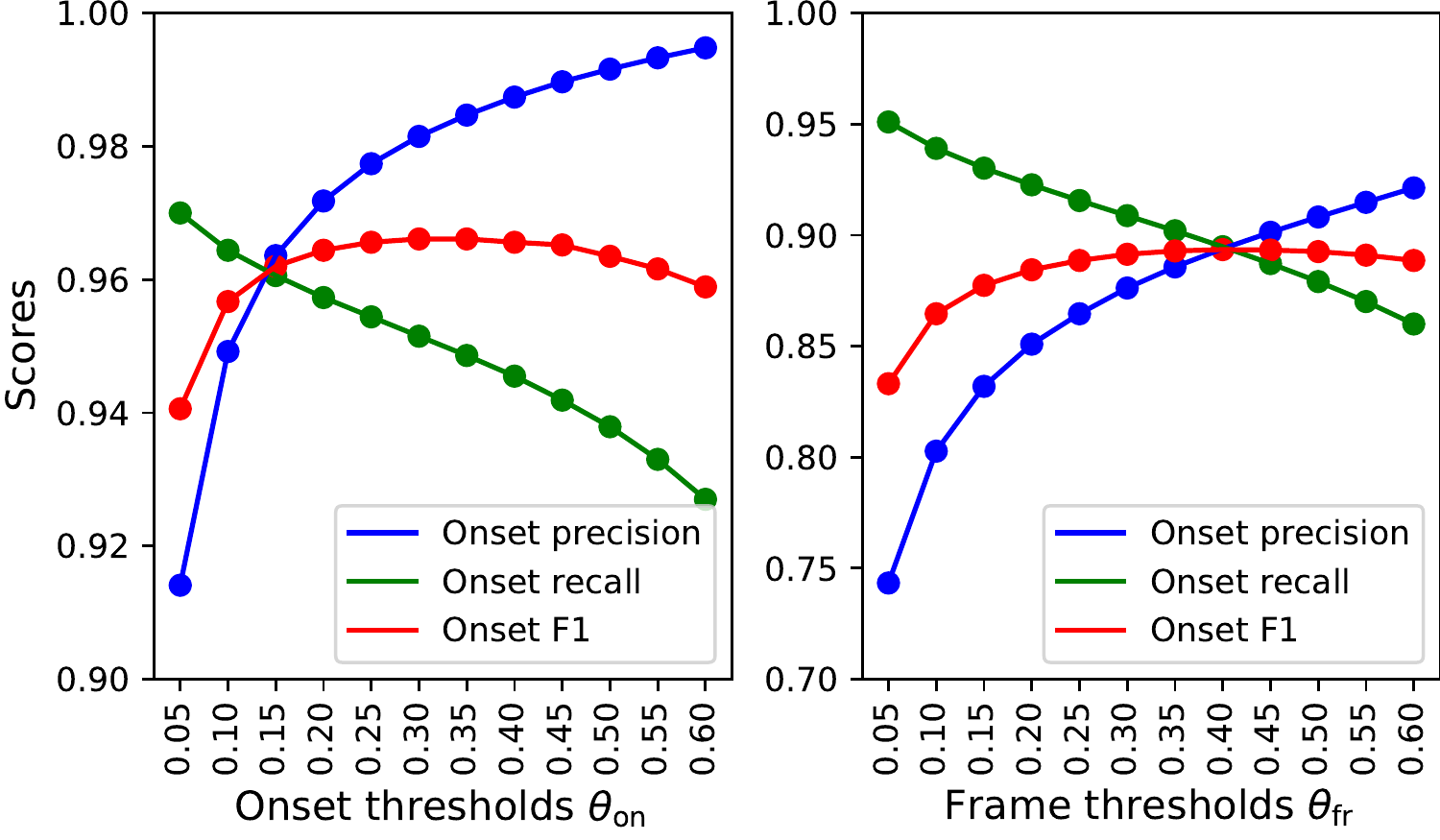}}
  \caption{Left: Onset-level precision, recall and F1 score evaluated with different onset thresholds $ \theta_{\text{on}} $. Right: Frame-level precision, recall and F1 score evaluated with different frame thresholds $ \theta_{\text{fr}} $.}
  \label{fig:thresholds}
\end{figure}

\begin{table*}
  \caption{Transcription results evaluated with misaligned labels}
  \vspace{6pt}
  \label{tab:noisy_labels}
  \centering
  \resizebox{\textwidth}{!}{%
  \begin{tabular}{lcccccccccccc}
    \toprule
    & \multicolumn{3}{c}{\textbf{\textsc{Frame}}} & \multicolumn{3}{c}{\textbf{\textsc{Note}}} & \multicolumn{3}{c}{\textbf{\textsc{Note w/ offset}}} & \multicolumn{3}{c}{\textbf{\textsc{Note w/ offset \& vel.}}} \\
	\cmidrule(lr){2-4} \cmidrule(lr){5-7} \cmidrule(lr){8-10} \cmidrule(lr){11-13} 
     & P (\%) & R (\%) & F1 (\%) & P (\%) & R (\%) & F1 (\%) & P (\%) & R (\%) & F1 (\%) & P (\%) & R (\%) & F1 (\%) \\
    \midrule
    Onsets \& frames (misaligned labels) & 80.93 & 90.93 & 85.54 & 65.59 & 93.06 & 76.52 & 44.40 & 63.36 & 51.92 & 40.41 & 57.64 & 47.25 \\
    Regress cond on. \& off. \& vel. (misaligned labels) & 84.65 & 91.36 & \textbf{87.79} & 98.65 & 94.30 & \textbf{96.39} & 80.59 & 77.09 & \textbf{78.77} & 77.35 & 74.02 & \textbf{75.62} \\
	\bottomrule
\end{tabular}}
\end{table*}

To evaluate different thresholds for piano transcription, we experiment with onset thresholds and frame thresholds from 0.05 to 0.60. Fig. \ref{fig:thresholds} shows the precision, recall, and F1 score of the piano note transcription system evaluated with different onset thresholds and offset thresholds. Fig. \ref{fig:thresholds} shows that F1 scores are similar when onset thresholds and frames thresholds are between 0.2 and 0.4. Higher thresholds lead to higher precision but lower recall. According to those observations, we set all thresholds to 0.3 in our work. Table \ref{tab:hyperparameter_J} shows the piano transcription results with different hyper-parameters $ J $. The evaluation scores of systems with different $ J $ are similar, indicating that $ J $ does not need to be tuned elaborately in our proposed system. Overall, setting $ J $ to 5 slightly outperforms other configurations.

To show that our proposed regression-based piano transcription system is robust to the misalignment of labels, we randomly shift the labels of onsets and offsets with a uniform distribution between $-A$ ms and $+A$ ms. To show that our system can generalize well to misaligned labels, we set $A$ to 50 ms which is the default onset tolerance for onset evaluation used by the mir\_eval toolkit \cite{raffel2014mir_eval}. The seventh row of Table \ref{tab:noisy_labels} shows that the note F1 score of \cite{hawthorne2017onsets} decreases from 93.92\% to 76.52\% when trained with misaligned labels. One explanation is that, the system \cite{hawthorne2017onsets} is sensitive to misaligned labels. On the other hand, the second row of Table \ref{tab:noisy_labels} shows that our proposed regression-based system achieves a note F1 of 96.39\%, compared to the system trained with correct labels of 96.72\%. Our system achieves an F1 of 75.62\% when evaluated with offsets and velocities, compared to the system trained with correct labels of 80.92\%. Those results indicate that our proposed system is robust to misaligned labels. 

\begin{figure}[t]
  \centering
  \centerline{\includegraphics[width=\columnwidth]{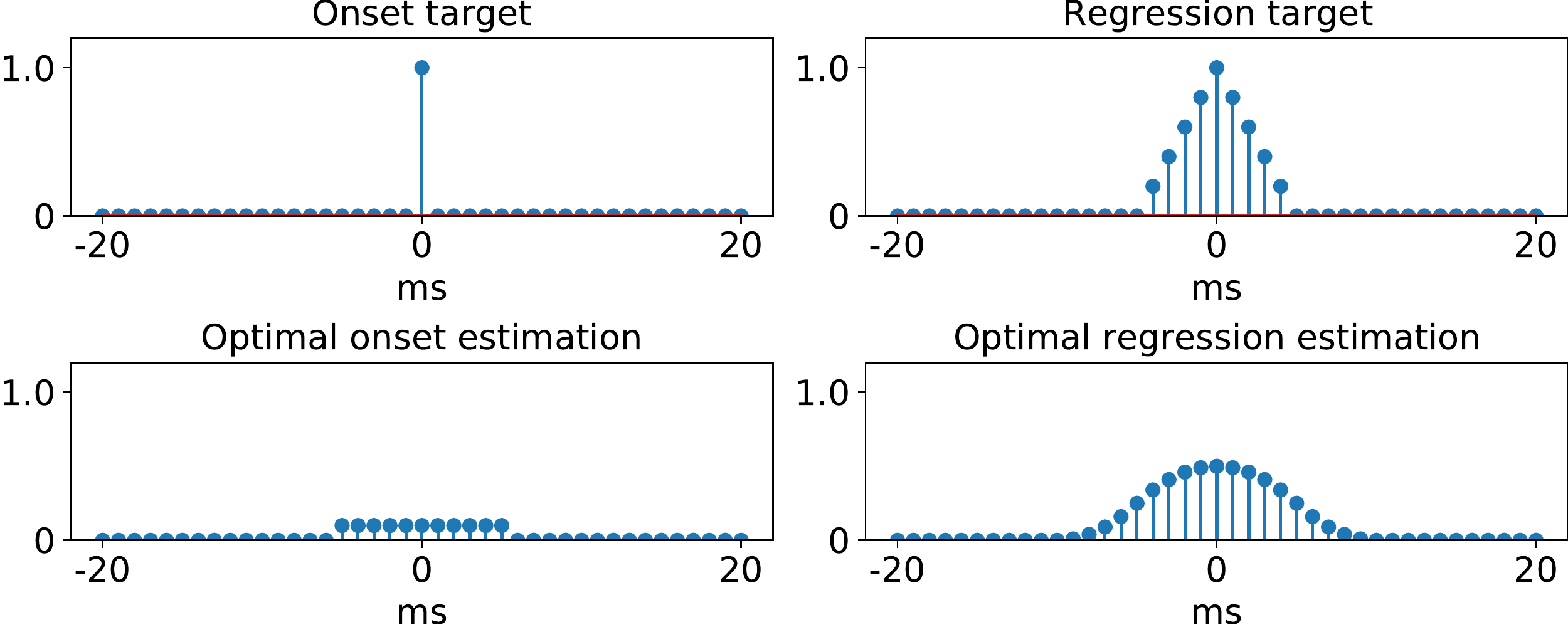}}
  \caption{Top left: onset target of \cite{hawthorne2017onsets}; Top right: our proprosed regression target; Bottom left: optimal onset estimation with misaligned labels; Bottom right: optimal regression estimation of misaligned labels.}
  \label{fig:noisy_labels}
\end{figure}

\qk{To mathematically demonstrate that our proposed regression targets are robust to misaligned labels, we denote the distribution of misaligned labels as $ q(t) $ and denote the target of a note as $ f(t) $, where $ f(t) $ can be either an onset target \cite{hawthorne2017onsets} shown in the top left of Fig. \ref{fig:noisy_labels} or our proposed regression target (\ref{eq:function_delta}) shown in the top right of Fig. \ref{fig:noisy_labels}. When using the binary cross-entropy loss function (\ref{eq:bce}) for training, the optimal estimation of targets is $ u(t) = f(t) \ast q(t) $ where the symbol $ \ast $ is a convolution operation. In absence of misaligned label, there is $ q(t) = \delta(t) $ and the optimal estimation $ u(t) $ is equivalent to $ f(t) $. When $ q(t) \sim [-A, A] $ is a uniform distribution, the optimal estimation $ u(t) $ of \cite{hawthorne2017onsets} and our proposed regression-based method are shown in the bottom row of Fig. \ref{fig:noisy_labels}. The bottom left of Fig. \ref{fig:noisy_labels} shows that the precise onset times are hard to obtain when using the targets of \cite{hawthorne2017onsets} in training. In contrast, the bottom right of Fig. \ref{fig:noisy_labels} shows that the precise onset time can be obtained whey using the regression-based method in training.}

\begin{figure}[t]
  \centering
  \centerline{\includegraphics[width=\columnwidth]{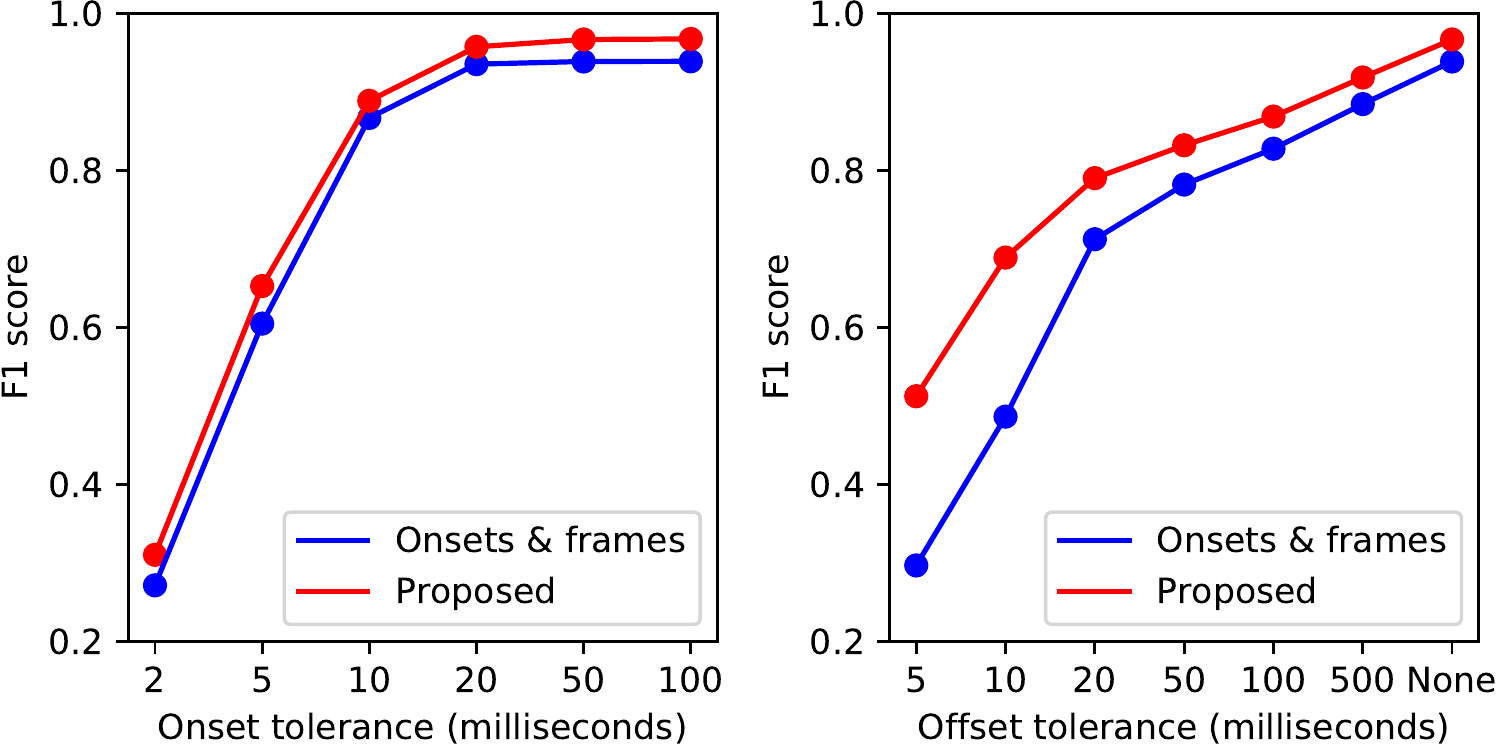}}
  \caption{Left: F1 score evaluated with various onset tolerances. Right: F1 score evaluated with various offset tolerances.}
  \label{fig:tolerances}
\end{figure}

\begin{table}
  \caption{Onsets evaluation with different onset tolerances}
  \vspace{6pt}
  \label{tab:onset_tolerance}
  \centering
  \resizebox{\columnwidth}{!}{%
  \begin{tabular}{lcccccc}
    \toprule
    & \multicolumn{3}{c}{\textbf{\textsc{Onsets \& frames}}} & \multicolumn{3}{c}{\textbf{\textsc{Proposed}}} \\
	\cmidrule(lr){2-4} \cmidrule(lr){5-7}
     & P (\%) & R (\%) & F1 (\%) & P (\%) & R (\%) & F1 (\%) \\
    \midrule
 100 ms & 99.54 & 89.24 & 93.94 & 98.24 & 95.42 & \textbf{96.79} \\
 50 ms & 99.52 & 89.23 & 93.92 & 89.17 & 95.35 & \textbf{96.72} \\
 20 ms & 99.14 & 88.92 & 93.58 & 97.22 & 94.45 & \textbf{95.79} \\
 10 ms & 91.74 & 92.53 & 86.73 & 90.19 & 87.69 & \textbf{88.91} \\
 5 ms & 63.89 & 57.69 & 60.53 & 66.22 & 64.47 & \textbf{65.32} \\
 2 ms & 28.69 & 25.93 & 27.19 & 31.49 & 30.68 & \textbf{31.08} \\   
	\bottomrule
\end{tabular}}
\end{table}

\begin{table}
  \caption{Onsets and offsets evaluation with different offset tolerances}
  \vspace{6pt}
  \label{tab:offset_tolerance}
  \centering
  \resizebox{\columnwidth}{!}{%
  \begin{tabular}{lcccccc}
    \toprule
    & \multicolumn{3}{c}{\textbf{\textsc{Onsets \& frames}}} & \multicolumn{3}{c}{\textbf{\textsc{Regress \& on. \& off.}}} \\
	\cmidrule(lr){2-4} \cmidrule(lr){5-7}
     & P (\%) & R (\%) & F1 (\%) & P (\%) & R (\%) & F1 (\%) \\
    \midrule
 500 ms & 93.71 & 84.13 & 88.49 & 93.25 & 90.59 & \textbf{91.88} \\
 200 ms & 87.65 & 78.75 & 82.81 & 88.19 & 85.68 & \textbf{86.90} \\
 100 ms & 82.80 & 74.41 & 78.24 & 84.48 & 82.08 & \textbf{83.25} \\
 50 ms & 75.47 & 67.76 & 71.28 & 80.24 & 77.95 & \textbf{79.06} \\
 20 ms & 51.66 & 46.19 & 48.68 & 69.96 & 67.97 & \textbf{68.94} \\
 10 ms & 31.64 & 28.14 & 29.73 & 52.04 & 50.57 & \textbf{51.28} \\
	\bottomrule
\end{tabular}}
\end{table}

\begin{table*}
  \caption{Pedal transcription evaluated on the test set of MAESTRO dataset.}
  \vspace{6pt}
  \label{tab:pedal}
  \centering
  \begin{tabular}{lccccccccc}
    \toprule
    & \multicolumn{3}{c}{\textbf{\textsc{Frame}}} &
    \multicolumn{3}{c}{\textbf{\textsc{Event}}} &
    \multicolumn{3}{c}{\textbf{\textsc{Event w/ offset}}} \\
	\cmidrule(lr){2-4} \cmidrule(lr){5-7} \cmidrule(lr){8-10} 
     & P (\%) & R (\%) & F1 (\%) & P (\%) & R (\%) & F1 (\%) & P (\%) & R (\%) & F1 (\%) \\
    \midrule
    Liang & 74.29 & 90.01 & 79.12 & - & - & - & - & - & - \\
    Onsets \& frames [our-implemented] & 94.30 & 94.42 & 94.25 & 93.20 & 90.26 & 91.57 & 86.94 & 84.28 & 85.47  \\
    Proposed & 94.30 & 94.42 & 94.25 & 91.59 & 92.41 & \textbf{91.86} & 86.36 & 87.02 & \textbf{86.58} \\
    \midrule
    Onsets \& frames (misaligned labels) & 93.62 & 94.14 & 93.77 & 92.71 & 85.48 & 88.69 & 83.17 & 77.03 & 79.78 \\
    proposed (misaligned labels) & 94.41 & 93.29 & 93.73 & 91.62 & 91.17 & \textbf{91.23} & 86.33 & 85.83 & \textbf{85.94} \\
	\bottomrule
\end{tabular}
\end{table*}

\begin{figure}[t]
  \centering
  \centerline{\includegraphics[width=\columnwidth]{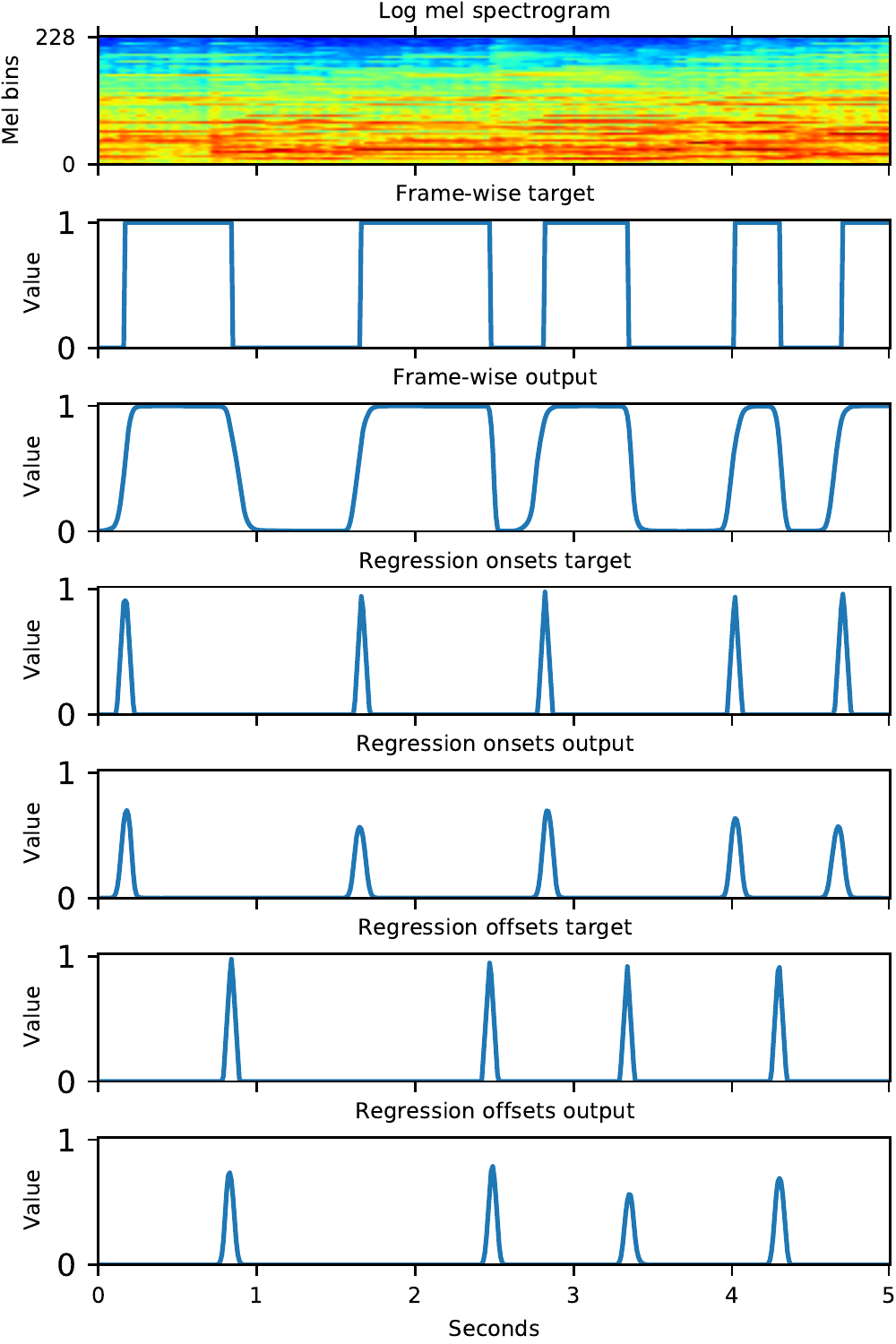}}
  \caption{Log mel spectrogram of a 5-second audio clip; pedal onset target; pedal onset output; pedal offset target; pedal offset output; pedal frame target; pedal frame output.}
  \label{fig:pedal_outputs}
\end{figure}

We evaluate piano transcription performance with different onset and offset tolerances that have not been evaluated in previous works \cite{hawthorne2017onsets, hawthorne2018enabling, kim2019adversarial}. The tolerances to be estimated range from 10 ms to 500 ms. Table \ref{tab:onset_tolerance} and the left part of Fig. \ref{fig:tolerances} show that with an onset tolerance of 2 ms, our system achieves an onset F1 score of 31.08\%. The F1 score increases to 88.91\% when onset tolerance is 10 ms, and increases to 96.79\% when onset tolerance is 100 ms. The F1 scores of our proposed system outperform the onsets and offsets system \cite{hawthorne2017onsets} in all onset tolerances. Table \ref{tab:offset_tolerance} and the right part of Fig. \ref{fig:tolerances} show the note F1 score evaluated with offset tolerances ranging from 10 ms to 500 ms when fixing onset tolerance to 50 ms. The \textit{None} label in Fig. \ref{fig:tolerances} 
equals to evaluation without offsets. Our system achieves a note F1 score of 51.28\% with an offset tolerance of 10 ms. The F1 score increases to 91.88\% with a tolerance of 500 ms. Our proposed system outperforms the onsets and frames system \cite{hawthorne2017onsets} in all offset tolerances. The experiments show that our proposed system can achieve higher transcription resolution than \cite{hawthorne2017onsets, hawthorne2018enabling}.

We evaluate the pedal transcription results as follows. Previous onsets and frames system \cite{hawthorne2017onsets} does not include sustain pedal transcription. Therefore, we implemented a pedal transcription with the onsets and frames system that has the same model architecture as our regression-based system for a fair comparison. In testing, for piano pieces without sustain pedals, the sustain pedal piano transcription system will produce no pedal events. We implemented the CNN-based method in \cite{liang2019piano}, and achieves a frame-wise precision, recall, and F1 score of 74.29\%, 90.01\%, and 79.12\%, respectively. Table \ref{tab:pedal} shows the pedal transcription result. Our system achieves an event-based F1 of 91.86\% evaluated with a pedal onset tolerance of 50 ms and achieves an event-based F1 of 86.58\% evaluated with both onset and offset tolerances of 50 ms and offset ratio of 0.2, outperforming our implemented onsets and frames system of 91.57\% and 85.47\% respectively. As far as we know, we are the first to evaluate sustain pedal transcription on the MAESTRO dataset. The first row of Fig. \ref{fig:pedal_outputs} shows the log mel spectrogram of the audio clip that is the same in Fig. \ref{fig:note_outputs}. The second and third rows show the frame-wise pedal target and system output. Values close to 1 indicate ``on'' states and values close to 0 indicate ``off'' states. The fourth and fifth rows show the regression onset targets and outputs. The sixth and seventh rows show the regression offset targets and outputs. Fig. \ref{fig:pedal_outputs} shows that our pedal transcription system performs well on the 5-second audio clip example.

\subsection{Error Analysis}
Error analysis was carried on a few transcribed pieces from real recordings. We observe that most false positives come from octave errors. Usually, the harmonics of notes are recognized as false positive notes. Other false positive errors include false positives with short durations of less than 15 milliseconds and false positive of repeated notes. There are also a small number of false positives that do not have an obvious explanation. For false negative (missing) errors, we observe that sometimes the higher notes of octaves can be missed. In addition, there are a small number of bass notes ignored by our system. We also observe degraded performance when a piano is out of tune or the qualities of recording devices are low.

\section{Conclusion}\label{section:conclusion}
We propose a high-resolution piano transcription system by regressing the precise onset and offset times of piano notes and pedals. At inference, we propose an analytical algorithm to calculate the precise onset and offset times. We show that our proposed system achieves a state-of-the-art onset F1 score of 96.72\% in piano note transcription, outperforming the onsets and frames system of 94.80\%. We show that our system is robust to the misaligned onset and offset labels. In addition, we investigate evaluating piano transcription systems with different onset and offset tolerances that were not evaluated in previous works. As far as we know, we are the first to evaluate pedal transcription on the MAESTRO dataset and achieves a pedal event F1 score of 91.86\%. One of the applications of our piano transcription system is the creation of the GiantMIDI-Piano dataset\footnote{\url{https://github.com/bytedance/GiantMIDI-Piano}}. Other applications include piano performance analysis, genre analysis, etc. The limitation of our proposed systems includes the transcription results depend on the quality of audio recordings and the transcription system need to be modified for real-time applications. In the future, we will investigate multi-instrument transcription using our proposed high-resolution transcription system.

\section{Acknowledgement}\label{section:acknowledgement}
We thank Mr. Mick Hamer for providing insightful analysis of our transcription system on several music pieces recorded on a Bosendorfer SE. We thank Prof. Gus Xia for providing a Yamaha Disklavier for playing back several transcribed MIDI files. We thank Mr. Hanying Feng for discussions on piano recording techniques.

\ifCLASSOPTIONcaptionsoff
  \newpage
\fi



%

\bibliographystyle{IEEEtran}
\bibliography{refs}

%





\end{document}